\documentclass{elsart}
\usepackage{times}
\usepackage[reqno]{amsmath}
\usepackage{amssymb}
\begin{document}
\begin{frontmatter}
\title{Pick-up ion dynamics at the structured quasi-perpendicular shock}
 \author{D. Zilbersher and M. Gedalin}
\address{Department of Physics, Ben-Gurion University, P.O. Box 653,
 Beer-Sheva, 84105, Israel}
 \begin{abstract} 
We study the pickup ion dynamics and mechanism of multiple reflection and 
acceleration at the structured quasi-perpendicular supercritical shock. The
 motion of the pickup 
ions in the shock is studied analytically and
numerically using the test particle analysis in the model shock front. The 
analysis shows that 
slow pickup ions may be accelerated at the shock ramp to high energies.
 The maximum ion energy is determined by the fine structure of 
the electro-magnetic field at the shock ramp and decreases when the angle
between magnetic field and shock normal decreases. Evolution of pickup ion
distribution across the nearly-perpendicular shock and pickup ion 
spectrum is also studied by direct numerical analysis. 
 \end{abstract}
\end{frontmatter}
%\psdraft
\section{Introduction}  
Interstellar pickup ions play an important role
in the physics of the outer heliosphere. They may modify both the 
large scale characteristics of the solar wind itself \cite{Hol72,Ise86}
and smaller scale solar wind structures such as interplanetary 
collisionless shock waves \cite{Zan95}. 
On the basis of the studies of the low-energy cosmic rays
it was proposed that the cosmic ray anomalous component originates from 
the interstellar pickup ions, accelerated at the  
quasi-perpendicular shocks \cite{Fis74,Jok86}. 

The observations, made
by Ulysses at 4.5AU \cite{Glo94}, have revealed accelerated interstellar 
pickup ions across the forward shock in the corotating interaction regions. 
It has been found that the injection 
efficiency for these pickup ions exceeds that one for solar wind ions, and 
that the accelerated pickup ions have power-law energetic spectra in the 
solar wind frame  \cite{Glo94}. Earlier investigations of 
the quasi-perpendicular cometary bow shocks have observed water-group pickup 
ions with speeds of several times the solar wind speed $v_u$ at and 
downstream of the shock\cite{Ipa86,Coa89}. 

Process of pickup ions production and features of the pickup ion 
distribution in 
the solar wind are studied quite well. Neutral atoms and molecules,  
penetrating from the local interstellar medium or escaping from comets, 
are ionized by photoionization, electron impact or charge exchange with 
the solar wind. Under influence of the solar wind electro-magnetic field 
the ions compose ring-beam distribution in velocity space. Then they are 
scattered by ambient and excited Alfvenic fluctuations 
to form spherical shell distribution, centered approximately 
at the solar wind velocity with a radius of about solar wind speed $V_u$
\cite{Sag86,LI87}.

Accelerated pickup ions were observed at quasi-perpendicular shocks, 
which are not able to 
accelerate incident thermal ions and pick up 
ions by the standard diffusive shock (Fermi) acceleration mechanism 
\cite{Web95}. The mechanism requires ions which can penetrate
the shock front in both directions. At quasi-perpendicular shocks it 
may occur when the ion 
velocity near the shock ramp is higher than a definite threshold speed, 
which significantly exceeds the solar wind speed \cite{Web95,Jok92}.
Hence, in order that pick up ions may be 
accelerated by the diffusive mechanisms, some preacceleration 
mechanism at quasi-perpendicular shocks must exist. 
In \cite{Zan95}, a detailed discussion was presented of
the following idea for the ion acceleration at a perpendicular shock
 proposed by \cite{Sag66}: if the ion 
encounters the shock with normal kinetic energy $m_iv_n^2/2$ much smaller 
than the
electrostatic potential $e\varphi_0$ at the shock, and the upstream Lorentz 
force is
directed toward the shock, then the ion finds itself trapped between the 
shock potential and the Lorentz force. Such ion is multiply reflected at 
the shock front and during each excursion to the upstream region it gains 
some 
energy until it is able to overcome the shock, when $m_iv_n^2/2>e \varphi_0$
or the Lorentz force in the normal direction exceeds the electrostatic 
force in the ramp. This idea    
for pickup ion preacceleration at the quasi-perpendicular shock has been 
developed further by several authors. Analytical estimates show 
that at the perpendicular shock the maximum energy 
gain for transmitted ions is proportional to the ratio of an solar wind 
ion gyroradius to the smallest characteristic scale of the electrostatic 
potential \cite{Zan95}. At the quasi-perpendicular shock maximum 
energy gain occurs when the ion escapes back into the upstream region 
\cite{Lee95}.
The multiple reflection ion acceleration gives very hard power law spectrum
\cite{Zan95}. The spectrum is weakly sensitive to the shock parameters and 
may extend to energy $\sim 0.5$MeV. Thus the injection problem associated 
with Fermi 
acceleration probably, may be solved by the multiple reflection ion 
acceleration mechanism.

In the present paper we study the details of the multiple reflected 
ion acceleration mechanism in the stationary structured quasi-perpendicular 
supercritical shock and determine the shock parameters, which control the 
pickup ion energy gain. We also consider evolution of the pickup ion 
distribution across the strong shock. Our approach differs from earlier 
ones in that we use field models qualitatively describing the actual
structure of the stationary fields at quasi-perpendicular super-critical 
shocks which 
consists of the extended foot, narrow ramp, overshoot and downstream region. 
In section 2 we consider analytically the pickup ion motion at the shock front
and derive the multiple reflection 
condition as a function of the field parameters. In section 3 we 
numerically analyze  the ion motion in the stationary model shock front. 
 The analysis illustrates the analytical 
consideration of multiple reflection process developed in section 2 and 
provides additional information about the process. The method also allows 
diagnostics of pickup ion distribution at the shock and of pickup ion 
spectra before the ramp and in the far upstream and downstream regions.

 \section{Pickup ion dynamics in the shock front}

When studying the pickup ion dynamics in the shock front we adopt the 
usual assumptions that the shock is one-dimensional and stationary. 
In doing so we do not consider the effects of, for example, 
 possible rippling of the 
shock surface  or interaction with waves, assuming that    the 
stationary electric and magnetic fields in the shock  front determine 
the ion behavior (see, however, discussion in sec. \ref{concl}).

We shall work in the normal incidence frame (N), where the upstream plasma 
velocity is along the shock normal. Let us choose the coordinates in 
such a way that the normal is along $x$ axis which is directed toward the 
downstream region (in \cite{Lee95} $x$ axis was directed toward the 
upstream region), upstream and downstream 
magnetic fields are in $xz$ plane, and the noncoplanarity direction 
 is along $y$ axis. Then the ion motion is governed by the following 
equations of motion:
\begin{align}
& m_i\dot{v}_x=e(E_x+v_yB_z-v_zB_y),\label{dvx}\\
& m_i\dot{v}_y=e(E_y+v_zB_x-v_xB_z), \label{dvy}\\
& m_i\dot{v}_z=e(v_xB_y-v_yB_x), \label{dvz}
\end{align}
where $B_y$, $B_z$, and $E_x=-d\phi/dx$ depend only on $x$, while 
$B_x=\text{const}$, and the motional electric field 
$E_y=V_uB_u\sin\theta=\text{const}$ (where $V_u$ is the upstream flow speed).
Here subscript $u$ denotes 
asymptotically homogeneous upstream parameters, and $\theta$ is the 
angle between the shock normal and upstream magnetic field vector. 

The qualitative profile of the magnetic field is known quite well. It 
is usually considered \cite{Sc86} to consist of the extended foot 
with $L_f\sim 0.5 
(V_u/\Omega_u)$ (where $\Omega_u=eB_u/m_i$ is the upstream ion 
gyrofrequency), on which the magnetic field $B_z$ gradually increases 
by the amount of $\lesssim B_u$. It is followed by the narrow ramp with 
the width $c/\omega_{pi}>L_r>c/\omega_{pe}$, where the main magnetic 
field $B_z$ jump occurs, and magnetic overshoot and probably large 
downstream magnetic field oscillations. The noncoplanar magnetic field component   
$B_y$ is always small relative to $B_z$, and negligible upstream. 
Substantial component of pickup ions may affect the shock profile and 
alter the typical scales (see detail discussion in \cite{Zan95}). Here we 
consider pickup ions as a low density test particle population.

The electric field profile is known much worse. It is distributed over 
the whole shock front including foot, ramp, and overshoot, and 
penetrates into the downstream region, so that only a part of the total 
cross-shock potential is applied at the ramp.The qualitative picture of 
the quasi-perpendicular shock front is shown in Figure~1. 

An ion with a low kinetic energy in $x$ direction 
($m_iv_x^2/2<<e \phi$) is unable to overcome the electrostatic 
potential at the ramp and is reflected back to the upstream region. 
If $v_yB_z>0$, upstream Lorentz force returns it to the ramp again.
In this way the ion becomes trapped near the ramp and 
quickly oscillates in $x$ direction until it escapes upstream or downstream 
region \cite{Zan95,Lee95}.

It does not seem possible to solve (and even analyze quantitatively) 
the ion equations of motion \eqref{dvx}-\eqref{dvz} in the general case 
of Figure~1. Instead we shall analyze them for the case of surfing 
ions, which are assumed to be trapped in the ramp vicinity. 
To do so we make an assumption (verified aposteriori) that these ions 
oscillate quickly in $x$ direction, while two other velocity 
components vary slowly on the oscillation period (such behavior can be 
expected since strong ion acceleration is possible only when $E_y$ acts on 
the ion for a substantially long time). In this case separating fast and 
slow motion, one has: 
\begin{align}
& m_i\dot{v}_x=-e\frac{d\phi_{\text{eff}}}{dx}, \label{dvx1}\\
& m_i\dot{v}_y=e(E_y+v_zB_x), \label{dvy1}\\
& m_i\dot{v}_z=-ev_yB_x, \label{dvz1}
\end{align}
where \eqref{dvx1} describes fast oscillations along $x$,  while 
\eqref{dvy1} and \eqref{dvz1} describe slow motion in $yz$ plane. The 
effective potential 
\begin{align}
&\phi_{\text{eff}}=\phi -v_yA_y-v_zA_z, \label{phieff}\\ 
& B_z=\frac{dA_y}{dx}, \qquad B_y=-\frac{dA_z}{dx}, 
\label{vectorpotential}
\end{align}
weakly depends on time via slow time-dependence of $v_y$ and $v_z$. 
Further simplification can be achieved by consideration a model 
profile, where ${\bf B}={\bf B}_u$ and ${\bf E}=(0,E_y,0)$ before the ramp, and  
Taylor expanding up to the first order 
\begin{align}
& B_y=(\frac{dB_y}{dx})|_{x=0}x, \label{byexp}\\
& B_z=B_u\sin\theta +(\frac{dB_z}{dx})|_{x=0}x, \label{bzexp}\\
& E_x=(\frac{dE_x}{dx})|_{x=0}x, \label{exexp}
\end{align}
in the vicinity $x>0$ of the upstream edge of the ramp ($x\rightarrow 
-\infty$ and $x\rightarrow 
+\infty$ corresponds to the asymptotically homogeneous upstream and 
downstream regions, respectively). This approach 
implicitly assumes that the trapped ions do not penetrate the ramp 
deeply. In this case the effective potential takes the following form
\begin{equation}\label{phieff1}
\phi_{\text{eff}}=\begin{cases}
-v_yB_u\sin\theta x, & x<0, \\
\tfrac{1}{2}\lambda x^2 -v_yB_u\sin\theta x, & x>0
\end{cases}
\end{equation}
where
\begin{equation}
\lambda=-\frac{dE_x}{dx}-v_y\frac{dB_z}{dx}+v_z\frac{dB_y}{dx}, 
\label{lambda}
\end{equation}  
where the derivatives are taken at $x=0$,
and $\phi_{\text{eff}}=0$ at the upstream edge of the 
ramp $x=0$. 

Estimating typical ${dE_x}/{dx}\sim \Delta \phi/L_r^2\sim 0.5 
m_iV_u^2/2eL_r^2$  and $dB_z/dx\sim \Delta B_z/L_r\sim B_u/ L_r$, where 
$L_r$ is the ramp width, one finds that the term ${dE_x}/{dx}$ in 
$\lambda$ dominates unless $v_y/V_u\gtrsim (V_u/\Omega_u)/L_r= 
M_A(c/\omega_{pi})/L_r$, where $M_A$ is Alfven Mach number. 

Equations \eqref{dvx1} and \eqref{phieff1} describe oscillations in a 
potential well (if $\lambda>0$, which is typical for the 
quasiperpendicular shock front), with the potential minimum of 
$\phi_{\text{min}}=-(v_yB_u\sin\theta)^2/2\lambda$ at 
$x=v_yB_u\sin\theta/\lambda$. The equations of motion are easily solved 
as follows. 

In the region $x<0$ one has (the initial condition is $v_x=v_{x0}$ - 
$x$-component of the reflected ion velocity before its excursion to the 
upstream region, $x=0$): 
\begin{equation}
v_x^2=v_{x0}^2+2v_y\Omega_ux, \label{vx2}
\end{equation}
while in the region $x>0$ the solution has the following form:
\begin{align}
&x=x_m+\Delta \sin(\omega t-\varphi_0), \label{xpos}\\
& v_x=\omega \Delta \cos(\omega t -\varphi_0), \label{vxpos}\\
& \varphi_0=\arcsin \frac{\beta}{(\beta^2+\lambda m_iv_{x0}^2/e)^{1/2}}, 
\label{phase}\\
& \omega=(e \lambda/m_i)^{1/2}, \quad \beta=v_yB_u, \quad x_m=\beta/\lambda, 
\quad \Delta=\frac{(\beta^2+\lambda m_iv_{x0}^2/e)^{1/2}}{\lambda}, 
\label{minmax}
\end{align}
where $v_{x0}$ is a slowly varying function of time.

When $v_y$ and $v_z$ vary slowly, the integral $\oint pdq=\oint 
m_iv_xdx$ over a closed trajectory is adiabatically invariant. Direct 
calculation gives
\begin{align}
&I=I_1+I_2, \quad I_1=\oint_{x<0}mv_xdx, \quad I_2=\oint_{x>0}mv_xdx, 
\label{adiabat1}\\
& I_1=\frac{1}{3}m_i^2\frac{v_{x0}^3}{\beta}, \label{adiabat2}\\
& I_2=\frac{m_i\omega}{\lambda^2} (\beta^2 
+\lambda m_iv_{x0}^2/e) [\frac{\pi}{2} + \arcsin\frac{\beta}{(\beta^2 + 
\lambda m_iv_{x0}^2/e)^{1/2}} \notag\\
&- \frac{1}{2}\frac{\lambda m_iv_{x0}}{(\beta^2 + 
\lambda m_iv_{x0}^2/e)^{1/2}}]=\text{const}. \label{adiabat3}
\end{align}    
It is easy to analyze the consequences in the limiting cases. If 
$\lambda m_iv_{x0}^2/\beta^2\ll 1$ (very low initial velocities at the 
upstream edge of the ramp), $I_2\gg I_1$ and
\begin{equation}
I_2\approx \frac{\pi \beta^2}{\lambda^{3/2}m_i^{1/2}}=\text{const}, 
\label{limit1}
\end{equation}
which corresponds actually to the case $v_{x0}\rightarrow 0$. In the 
opposite limit $\lambda m_iv_0^2/\beta^2\gg 1$ the upstream part 
dominates $I_1\gg I_2$ and 
\begin{equation}
I_1\approx \frac{m_i^2v_{x0}^3}{\beta}=\text{const}, \label{limit2}
\end{equation}
which immediately gives $v_{x0}\propto v_y^{1/3}$ \cite{Lee95}. 

Maximum kinetic energy of the oscillations is determined by the energy 
conservation:
\begin{equation}
(\frac{m_iv_x^2}{2})_{\text{max}}= \frac{m_iv_{x0}^2}{2} + 
(v_yB_u\sin\theta)^2/2\lambda, \label{vxmax}
\end{equation}
and remains relatively small with the dependence weaker than 
$v_x\propto v_y$.                                

Roughly estimating, the ion trapping ceases and it is transmitted 
downstream if it crosses the middle of the ramp. Hence the additional 
trapping condition is $m_iv_{x0}^2/2<e\phi(x=L_r/2)$, or
\begin{equation}
\frac{m_iv_{0}^2}{2}(\frac{v_y}{v_{y0}})^{2/3} + \frac{v_y B_u 
L_r}{2} < \frac{e\phi}{2}, \label{condition2}
\end{equation}
where $v_{0}$ is the initial ion $x$ velocity, $v_{y0}$ is its initial 
$y$ velocity, and $\phi$ is the cross-ramp potential. Taking into 
account that $v_0$ is small and assuming that $dE_x/dx$ is the dominant 
contribution in $\lambda$, one finds the following estimate for the 
maximum $v_y$ during the trapping:
\begin{equation} 
 (v_y)_{\text{max}}^{(\text{trap})}
 =\frac{e\phi}{B_uL_r} -\frac{m_iv_0^2}{B_uL_r} 
 \left(\frac{e\phi}{B_uL_rv_{y0}}\right)^{1/3}. \label{estimate}
 \end{equation} 

The second term in the last equation is a correction to the step 
function based estimate for $(v_y)_{\text{max}}$  \cite{Zan95}.
 
On the other hand, the solution of the slow equations with the initial 
condition $v_y=v_{y0}$, $v_z=v_{z0}$ is
\begin{eqnarray}
&& v_y=v_{y0}\cos(\Omega_u\cos\theta t) +(v_{z0}+V_u\tan\theta) \sin( 
\Omega_u\cos\theta t), \label{vxslow}\\
&& v_z=(v_{z0}+ V_u\tan\theta) \cos(\Omega_u\cos\theta t) - v_{y0} 
\sin(\Omega_u\cos\theta t) -V_u\tan\theta,  \label{vyslow}
\end{eqnarray}
and 
\begin{equation}
(v_y)_{\text{max}}^{(\text{slow})}=[v_{y0}^2+(v_{z0}+V_u\tan\theta)^2]^{1/2}. 
\label{estimate1}
\end{equation}
If $(v_y)_{\text{max}}^{(\text{slow})} < 
(v_y)_{\text{max}}^{(\text{trap})}$, the ion is not transmitted 
downstream but remains trapped and its $v_y$ and $v_x$ decrease until 
it escapes upstream. 

It is difficult to determine precisely the moment when the ion escapes 
upstream. We shall estimate the escape conditions assuming that the 
escape itself occurs when $|v_x|$ and $v_y$ come back (decrease again) 
to their initial values. In that case $v_z$ can be easily determined 
using energy conservation in the de Hoffman-Teller frame (HT, in which the 
plasma upstream velocity is along the upstream magnetic field). The 
transformation rule between the two frames is
\begin{align}
& v_x^{(N)}=v_x^{(HT)}, \quad 
v_y^{(N)}=v_y^{(HT)},\label{transform1}\\
& v_z^{(N)}=v_z^{(HT)}-V_u\tan\theta. \label{transform2}
\end{align}
Since $E_x$  is negligible upstream, the total HT potential is zero in 
this region and energy conservation gives $({\bf 
v}^{(HT)})^2=\text{const}$, which in our case results in the 
following estimate of the escape velocity:
\begin{equation}
 v_z^{(N,\text{escape})}=-v_{z0}^{(N)}-2V_u\tan\theta. \label{escapez}
 \end{equation} 
 It is worth mentioning that in this picture the ions experience 
 reflection in which their noncoplanar velocity $v_x$ does not 
 change, while others ($v_x$ and $v_z$) change their sign in HT. 
 
 It should be mentioned that the above analysis is approximate (since 
 the shock profile is treated in a specific model) and applies only to those 
 ions which are trapped and subsequently either transmit downstream and 
 do not appear at the ramp anymore or escape upstream. The analysis is 
 not able to catch  trajectories which do not meet these assumptions 
 and we have to consider them numerically in the next section. We also 
 do not consider here the ions which experience the shock drift 
 acceleration (cf. \cite{Zan95}).  

\section{Numerical analysis.}

In order to illustrate the above theoretical analysis and  
study the 
features, which cannot be studied analytically, we perform a test 
particle numerical analysis of pickup ion trajectories in a model shock 
front, which is taken to resemble the observed shock profiles, as in 
Figure~1. The corresponding model analytical form for the fields is 
taken as proposed by \cite{Ged95}: 
\begin{align}
&B_x=B_u \cos \theta=\text{const}, \label{bx} \\
&B_y(x)=\frac{cB_x}{4 \pi neV_u} \frac{dB_z(x)}{dx}, \label{by} \\
&B_z(x)=B_u \sin \theta(1+ \notag \\ 
&\frac {R_f-1}{2} (1+\tanh\frac{3(x+D_f-3D_r)}{D_f})
+\frac{R_r-R_f}{2} (1+\tanh \frac {3x}{D_r}) \notag \\
&+(R_o-R_r) \exp (- \frac{2(x-D_o)^2}{D_o^2}) \notag  \\
&+ \frac {R_d-R_r}{2} (1+\tanh \frac{3(x-D_o-D_d)}{D_d}), \label{bz} \\
&E_x(x)=- \frac{1}{ne} \frac{dp_{e,xx}}{dx}- \frac{1}{8 \pi ne}
\frac{dB_z^2}{dx} - \frac{\phi}{\sqrt{\pi}d_e} \exp(-\frac {x^2}{d_e^2})
\label{ex} \\
&E_y=\frac{{\bf V_u \times B_u}}{c}=\text{const},  E_z=0. \label{ey}
\end{align}
where $d_e$ is the scale of the additional Gaussian electric field 
variation, $D_f$, $D_r$ and $D_o$ control the foot, ramp and overshoot 
thicknesses respectively, 
$p_{e,xx} \propto n^{\gamma_e}$, $n \sim B_z$. The suggested form 
of $B_z$ describes qualitatively the front structure: foot at 
$-D_f<x<-D_r/2$, 
ramp $-D_r/2<x<D_r/2$, and overshoot $D_r/2<x<(D_o+D_d)$ (it should be noted 
that here for convenience $x=0$ corresponds to the middle of the ramp). 
The noncoplanar component $B_y$ is in agreement with the 
results of \cite{JE87,Ged96}. The suggested analytical profile  of the 
cross-shock electric field consists of the well-known hydrodynamical 
part (first two terms) for polytropic electrons, and additional 
Gaussian to take into account the penetration of the field in deep 
downstream and foot and provide the observed values of the total 
cross-shock potential drop.     
For the specific needs of the present paper the parameters were chosen 
as follows: $R_f=1.5$, $R_r=5$, 
$R_o=6$, $R_d=3.2$, $D_f=0.3(V_u/\Omega_u)$, 
$D_r=0.01(V_u/\Omega_u)$ (this parameter was varied, see below), 
$D_o=0.1(V_u/\Omega_u)$, $D_d=0.15(V_u/\Omega_u)$, and $\gamma_e=2$. 
We  chose to examine a high-Mach number  $M_A=7.5$.

The initial pickup ion distribution was assumed to be the following shell 
distribution \cite[]{Lie93,Gia95} 
\begin{equation}
f({\bf v})=\frac{n_0}{4 \pi V_u^2} \delta(|{\bf v}-{\bf V_u}|-V_u).
\label{f0}
\end{equation}
at the upstream edge of the foot.

Pickup ion trajectories are traced through the described shock front 
starting at the upstream edge of the foot $x=-D_f$. Figure~2 
presents the 
ion trajectories projected onto the $xy$-plane at the quasi-perpendicular 
shock with $\theta=70^\circ$. A variety of pickup ion 
trajectories are seen: 1) ions which are directly transmitted 
downstream, 
2) ions which cross the ramp from downstream to upstream once or 
several times, 3) ions which are reflected in the ramp only once, 4) ions 
which are 
accelerated by multiple reflection mechanism and then escape either 
to downstream or to upstream, and 5) multiply reflected ions which rotate 
around the ramp until they escape upstream.

Next several figures provide more qualitative details about 
the multiple 
reflection (trapping) process.Figure 3
shows the trajectory of an ion which is transmitted downstream after 
several reflections. It is seen that at 
each next encounter with the 
ramp the ion penetrates into the ramp more deeply, while both $v_x$ 
and $v_y$ increase, until the ion crosses the ramp at the maximum of 
$v_y$ and drifts downstream having a high gyration velocity. The 
initial ion energy is about $m_iV_u^2/2$, and it is accelerated to 
about four times higher 
downstream energy in four cycles  of reflection. 

Figure~4 shows 
the trajectory of an ion which escapes upstream after 
making a full cycle of multiple reflections, during which $v_y$ 
increases and decreases again. Figure~4a shows that the ion trajectory does 
not cross the middle of the ramp. From Figure~4b one can see that the 
amplitude of oscillations of $v_x$ is roughly proportional to $v_y$, 
that is, the adiabatic approximation works quite well. It is seen also 
that the reflection-escape process is indeed almost specular in 
$v_x-v_y$ plane: $v_y$ is the same at the entry and escape points, 
while $v_x$ changes its sign. As can be seen from Figure~4c, $v_z$ 
in the last escape point closely corresponds to \eqref{escapez}.

Figure~5 shows
the trajectory which is not covered by the above theoretical 
analysis.  The ion is trapped near the ramp for some time and 
afterwards it becomes trapped around the ramp, making several large 
amplitude gyrations and crossing the shock front back and forth. 
Eventually it escapes upstream (high negative final $v_z$ in 
Figure~5c) having a substantial gyration velocity.  

Figure~6a 
shows which part of the initial pickup ion distribution undergoes multiple 
reflection (for perpendicular shock geometry). 
In Figure~6b we show the same distribution of incident 
pickup ions at the upstream edge of the ramp, where it is already 
strongly disturbed. Almost all multiply reflected ions are taken from 
the low $v_x$ part of the distribution, in agreement with previous 
theoretical observations \cite{Zan95,Lee95}. Most of them have substantial 
positive $v_y$  at their entry to the ramp, as is expected according to the 
analytical consideration.  

The next several figures present the evolution of the pickup ion 
distribution  across the structured shock front.
In Figure~7
we present the pickup ion distributions obtained numerically by 
tracing the initial shell distribution \eqref{f0} across the 
perpendicular shock. The distributions are plotted for several 
positions: (a) before the foot at $x=-0.6(V_u/\Omega_u)$, (b) at the 
upstream edge of the ramp $x=-0.005(V_u/\Omega_u)$, (c) at the 
downstream edge of the ramp  $x=0.005(V_u/\Omega_u)$,  
and (d) far downstream at $x=5(V_u/\Omega_u)$. 
The ion distribution before the foot (Figure~7a) consists of the 
incident pickup ion shell and ions which are reflected in the way 
similar to the reflection of the ions from the wings of the thermal ion 
distribution \cite{Ged95}. The ion distribution at the upstream edge 
of the ramp (Figure~7b) includes these reflected ions and much more 
energetic trapped accelerated ions, which are seen also just behind the 
ramp in Figure~7c (only those which crossed the ramp and are 
transmitted further downstream). Far downstream distribution 
(Figure~7d) consists of low energy reflected ions and high energy 
multiply reflected ones. Since there is not gyrophase mixing in the 
perpendicular shock (all ions have the same downstream drift 
velocity) the downstream distribution is spatially dependent (actually 
periodic). 

Figure~8 
present the evolution of pick up ion distribution 
across the  nearly perpendicular $\theta=80^\circ$ 
shock. For each position both $v_x,v_y$ and $v_y,v_z$ projections are 
shown.
The distribution 
before the foot (Figures~8a and 8b) differs from that one in the 
perpendicular case (Figure~7a) only by presence of ions which escaped 
upstream (high negative $v_z$). The upstream edge distribution  
(Figures~8c and 8d) also 
shows presence of these escaping ions. The multiply reflected ions in 
Figure~8d lie on two semicircles, corresponding to the slow rotation in 
$v_y,v_z$ plane (see \eqref{dvy1} and \eqref{dvz1}), separately for positive and 
negative initial $v_z$.   
The most remarkable difference from the perpendicular case can be seen 
in the downstream distribution (Figures~8g and 8h), which shows strong 
phase mixing. The downstream accelerated pickup ions are situated on 
two (almost) hemispheres, corresponding to the rotation of the 
semi-circles in Figure~8f.

The numerical analysis allows also to obtain the differential energy 
spectra of accelerated ions. Figure~9
 shows the far upstream 
(Figure~9a) and far downstream (Figure~9b) ion energy spectra $dN/d\epsilon$,
\begin{equation}
\frac{dN}{d\epsilon}=\frac{m_iV_u^2}{2} \frac{v}{m_i} \int 
f(v,\theta,\varphi) \sin \theta d\theta d\varphi
\end{equation}
where $\epsilon$ is the dimensionless ion energy, 
$\epsilon=(m_iv^2/2)/(m_iV_u^2/2)$. The far upstream  spectrum consists 
of the dense population of incident ions 
($dN/d\epsilon=\text{const}$) and low density high energy  component 
of accelerated 
ions which escaped upstream for two different angles between the shock 
normal: $\theta=80^\circ$ (dotted line) and $\theta=70^\circ$ (dashed 
line). With the decrease of the angle the number of escape ions 
(Figure~9a)
increases but their energy decreases.  One can see also the drastic 
drop of acceleration efficiency with the increase of obliquity 
in the downstream distribution of 
pickup ions, where the highest energy of accelerated ions drops from 
$10^2$ (in $m_iV_u^2/2$) to slightly higher than $10^1$ when the angle 
decreases from $80^\circ$ to $70^\circ$.

Figure~10 
shows the accelerated pickup ion distribution just before 
the ramp and far downstream for two values of the total cross-shock 
potential. As could be expected, the acceleration efficiency decreases 
with the decrease of the potential, although this decrease is almost 
not noticeable in the far downstream distribution because of the 
logarithmic scale. It is clearly seen that the distribution is nearly 
exponential $dN/d\epsilon\propto \exp (-\alpha \epsilon)$ at the upstream 
edge of the ramp, with $\alpha\approx 0.020$ for the potential 
$\varphi=0.7(m_iV_u^2/2)$, and  $\alpha\approx 0.033$
 for $\varphi=0.5(m_iV_u^2/2)$. The downstream distributions
reveal power spectra $dN/d\epsilon\propto \epsilon^{-\beta}$, where 
$\beta \approx 3/2$ and is almost independent of the cross-shock 
potential. The result is in conformity with the power low tail in energy 
produced by multiply reflected ion acceleration at strong perpendicular 
shock found in \cite{Zan95}.

Dependence of the accelerated ion spectra on the ramp width is shown 
in Figure~11, 
where the upstream-ramp-edge  and far downstream 
spectra are presented for the ramp width $D_r=0.01(V_u/\Omega_u)$ 
(solid line) and $D_r=0.03(V_u/\Omega_u)$ (dotted line). As is 
expected, the acceleration efficiency drops drastically when the sock 
becomes wider. It could be expected, however, that any substructure in 
the ramp would enhance the acceleration, as can be seen from 
Figure~12,
 where the ion spectra are compared for the case of 
$D_r=0.03(V_u/\Omega_u)$ without (dotted line) and with a substructure 
(dashed line). Presence of such internal substructure plays the role 
of a narrow ramp.

It is worth to mention that our approach allows also to study the behavior of 
heavy ions. Figure~13 
presents the results of such analysis, comparing 
the downstream spectra of ions with the mass $m_i=4m_p$ (which would correspond 
to  He$^+$ ions) and $m_i=15m_p$ (O$^+$). The 
initial distribution is the same pickup ion shell as for the case 
$m_i=m_p$, considered throughout the paper. It can be seen that the 
acceleration is less efficient for heavy ions than for protons
 (Figures~9b and~10b). Detailed analysis of the 
heavy ion behavior is beyond the scope of the present paper and will 
be presented elsewhere.

\section{Conclusions}\label{concl}

We have considered the surfing mechanism of the pickup ion energization 
at  strong high-Mach number 
quasi-perpendicular shocks. The  distribution of the electro-magnetic 
field at the shock front determines the ion motion across the shock,and 
the ion reflection and acceleration processes depend on the details of 
 the fine 
structure of the shock front. We had to make a choice of the shock 
structure in order to determine quantitatively the features of the 
accelerated ions and their dependence on the shock parameters.
It is clear that a deviation of the actual shock
structure from the model adopted here would give somewhat different 
quantitative results, which cannot be predicted unless we know profiles of 
the fields at the shock. However, some tendencies can be predicted on the 
basis of the analytical and numerical investigations represented in 
sections 2 and 3.

The conditions for the multiple ion reflection  and the maximum energy 
gain are 
determined by the slope of the electric and magnetic field profiles at the 
ramp and therefore very sensitive to the ramp width. 
For the chosen model the scales of the magnetic and electric field 
variations are the same and $\approx D_r$. The analytical 
consideration gives the following estimate for the maximum downstream 
ion energy in the nearly perpendicular shock
(see also \cite{Lee95,Zan95}):
\begin{equation}
{\mathcal E}_{\text{max}} \approx \frac{m_iV_u^2}{2}(\frac{2\phi}{D_rR})^2, 
\label{max}
\end{equation}
where $\phi=e\varphi/(m_iV_u^2/2)$, $R=B_{zr}/B_u$ (where $B_{zr}$ is $z$ 
component of the magnetic field at the downstream edge of the ramp), and 
$D_r$ is 
measured in $V_u/\Omega_u$. For the initial ion energy of ${\mathcal E}=1 
\text{keV}$, ramp width of $D_r\approx 0.01(V_u/\Omega_u)$,  
cross-shock potential $\phi=0.5$, and magnetic compression ratio 
$R=2.5$, one finds ${\mathcal E}_{\text{max}}\approx 1\text{MeV}$, which is 
probably sufficient for injection into diffusive acceleration regime.    

Since the efficiency is $\propto D_r^{-2}$ and decreases with the 
increase of obliquity, 
the pick up ion injection for Fermi acceleration mechanism seems to be 
effective 
only at nearly-perpendicular shocks waves with very narrow ramp or when 
some substructure is present which reduced the effective width of the 
ramp. 

The dependence of the acceleration features on the Mach number is 
determined mostly by the dependence of the shock width and 
magnetic compression ratio on the Mach number. 
In the absence of a satisfactory theory, 
which could provide these dependencies, the only conclusion is that 
the mechanism efficiency should rapidly decrease with the decrease of 
the Mach number because of the shock widening.

The obtained model downstream accelerated ion spectra $\propto 
\epsilon^{-3/2}$ (see also \cite{Zan95}), which corresponds to $f(v)\propto 
v^{-2}$ (and is in agreement with earlier studies), is harder 
than $v^{-4}$ that is observed. This discrepancy may be probably 
attributed to the mentioned deviation of the actual shock structure 
from the adopted model. 

It should be mentioned that, due to the high sensitivity of the 
trapping-detrapping mechanisms to the details of the magnetic and 
electric fields in the ramp, it could be in principle affected by even 
relatively weak deviations from stationarity (for example, presence of 
large amplitude waves in the ramp) or one-dimensionality (like 
rippling of the shock surface), as well as presence of any 
small-scale substructure inside the ramp. Observation evidence is not 
unambiguous, and  the effect of such  deviations from the 
model is not clear apriori and requires special study which is 
beyond the scope of the present paper.  

In summary, the analysis allowed to shed additional light on the 
importance of the small-scale structure of the shock front (in 
particular, high gradients of the electric field) for the ion 
acceleration processes. The test particle numerical analysis allowed 
to determine the details of ion 
behavior and the spatial and energetical distribution of accelerated 
ions in the shock front.

%\acknowledgements
This research was partially supported by grant  94-00047 from the United 
States-Israel Binational Science Foundation (BSF), Jerusalem, Israel 
and partially  by THE ISRAEL SCIENCE FOUNDATION founded by The Israel 
Academy of Sciences and Humanities.

\end{document}